\documentclass[twocolumn,aps,prd, floatfix,preprintnumbers,superscriptaddress,showpacs]{revtex4-1}
\usepackage{graphicx}
\usepackage{slashed,color}
\usepackage{amsmath,amsfonts,bm}
\usepackage{amssymb}
\usepackage{wasysym}
\usepackage{hyperref}

\newcommand{\nn}{\nonumber \\ }

\hypersetup{colorlinks, linkcolor = [rgb]{0,0.0,1.0}, citecolor = [rgb]{0,0.0,1.0}, urlcolor = [rgb]{0,0.0,1.0}}

\begin{document}



\author{A.~V.~Radyushkin}
\affiliation{Old Dominion University, Norfolk,
             VA 23529, USA}
\affiliation{Thomas Jefferson National Accelerator Facility,
              Newport News, VA 23606, USA
}

\title{Double  Distributions and Pseudo-Distributions}

\begin{abstract}
    We describe   the  approach to  lattice extraction of    Generalized Parton Distributions (GPDs)  that is 
     based on the use
    of the double distributions (DDs) formalism within the  pseudo-distribution framework. 
    The advantage of using   DDs is that GPDs obtained in this way  have the mandatory   polynomiality 
    property,   a non-trivial correlation between $x$- and $\xi$-dependences of GPDs. 
   Another   advantage of using DDs is  that the $D$-term appears  as an independent entity
   in the DD formalism
 rather than   a part of GPDs $H$ and $E$. 
We   relate the $\xi$-dependence of GPDs to  the width of the $\alpha$-profiles of the corresponding 
DDs, and discuss  strategies for fitting lattice-extracted pseudo-distributions by DDs.  
The approach described in the present paper may be  used  in ongoing and future  lattice extractions of GPDs.

\end{abstract}

\maketitle


\section{
 Introduction}

  Generalized Parton Distributions (GPDs) \cite{Mueller:1998fv,Ji:1996ek,Radyushkin:1996nd,Radyushkin:1996ru,Ji:1996nm,Radyushkin:1997ki}
(for reviews see \cite{Ji:1998pc, Diehl:2003ny,Belitsky:2005qn})
are a major object of study at  future Electron-Ion Collider and existing facilities at Jefferson Lab and CERN.
They provide a detailed information about hadronic structure. Being functions $H(x,\xi,t)$  of 3 kinematic variables
  (while there are other  GPDs: $E, \tilde H, \tilde E$, etc.,  we will use  $H$ as a generic notation), 
they combine properties of usual parton distributions $f(x)$, 
hadronic form factors $F(t)$ and, in the central region $|x|< \xi$, 
of the  distribution 
amplitudes $\varphi (x/\xi)$.

However, this multi-dimensional nature of GPDs highly complicates their extraction from 
experimental data. In particular,  deeply virtual Compton scattering (DVCS), which is the main  tool for obtaining information  about GPDs, gives information about GPDs on the lines $x= \pm \xi$ or 
through the Compton form factors that are $x$-integrals of GPDs with the $1/(x-\xi)$ weight.

More complicated processes like  double DVCS or recently proposed single diffractive hard exclusive
photoproduction 
 \cite{Qiu:2023mrm} 
may provide information about GPDs off the $x=\pm \xi$ diagonals. 
The study of such processes is  in its  early stage. 

During the last decade, starting with the pioneering paper of X. Ji \cite{Ji:2013dva} that introduced 
the quasi-distribution approach (see also Ref. \cite{Ma:2014jla} for  ``lattice cross sections'' approach), 
strong efforts have been made to calculate parton distributions on the lattice
(for reviews see Refs. \cite{Cichy:2018mum,Constantinou:2020pek,Ji:2020ect,Constantinou:2020hdm}).   In particular, matching conditions for  GPDs in  the quasi-distribution 
approach were discussed in Refs. \cite{Ji:2015qla,Xiong:2015nua,Liu:2019urm}.
For a review of   recent  lattice calculations  of GPDs see  Refs. \cite{Lin:2023kxn,Cichy:2023dgk}.

In our  paper \cite{Radyushkin:2019owq},   general aspects  of lattice QCD extraction of GPDs have been discussed
in the framework of the pseudo-distribution approach \cite{Radyushkin:2017cyf,Radyushkin:2019mye}. 
The advantage of  lattice calculations is that  matrix elements $M(\nu, \xi, t)$ (``Ioffe-time distributions'' (ITDs))
of nonlocal operators measured 
on  the lattice are related to  Fourier transforms of GPDs  $H(x,\xi,t)$,  which may be inverted using
various technics to produce  GPDs as functions of $x$ for fixed values of skewness $\xi$ and 
invariant momentum transfer $t$. 

An important property of GPDs  is {\it polynomiality} \cite{Ji:1998pc}, which states that $x^N$ moment of $H(x,\xi,t)$ 
must be a polynomial of $\xi$ of not larger than $(N+1)$th  power. This nontrivial correlation 
between $x$- and $\xi$- dependences of $H(x,\xi,t)$  is automatically  satisfied when GPDs 
are obtained from double distributions   $F(\beta, \alpha,t)$ \cite{Mueller:1998fv,Radyushkin:1996nd,Radyushkin:1996ru,Radyushkin:1998es,Radyushkin:1998bz}. 

 The goal of the present work is to outline  the  approach of lattice extraction of double distributions 
 from lattice calculations. The paper organized as follows.
 To make it self-contained, in Sec. II we formulate the definitions of usual (light-cone)  GPDs,  DDs and discuss their relationship. 
 Some basic properties of GPDs are discussed in Sec III.  There we also introduce Ioffe-time distributions.
 Pseudo-distributions, as generalizations of the ITDs onto  correlators  off the light cone are introduced 
 in Sec. IV.  Some strategies for fitting lattice-extracted pseudo-distributions by DDs are discussed in Sec. V. 
 Finally, in  Section VI, we summarize our discussion. 


 \setcounter{equation}{0}  

     \setcounter{equation}{0}
 
 \section{GPDs and DDs} 
 
\subsection{Definition of GPD }

\begin{figure}[htb]
\centerline{\includegraphics[height=3.8cm]{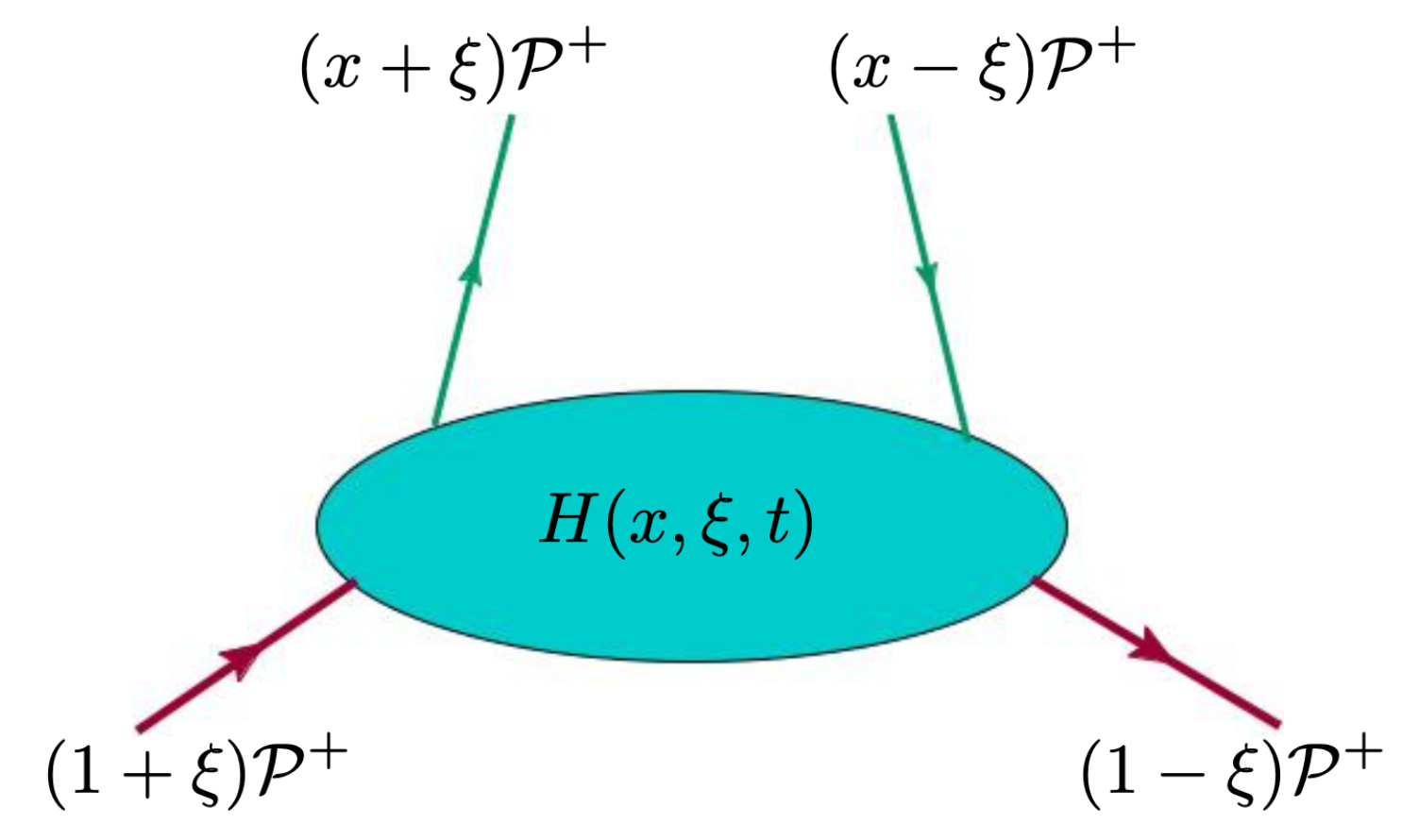}}
\caption{Flux of the momentum plus-components in terms of GPD variables.}
\label{FluxGPD}
\end{figure}

In the GPD description  of a nonforward kinematics   proposed by X. Ji \cite{Ji:1996ek}, 
the plus-components of 
the initial $p$  and final $p'$  hadron momenta
are given by  $(1+\xi){\cal P}^+$ and  $(1-\xi){\cal P}^+$, respectively, with 
${\cal P}$ being the  average  momentum ${\cal P}= (p+p')/2$, while the partons
have $(x+\xi){\cal P}^+$ and $(x-\xi){\cal P}^+$ as the plus-components of their momenta, see Fig.~\ref{FluxGPD}.

For the pion, one may define the   light-cone GPDs  $H(x,\xi,t;\mu^2)$  \cite{Ji:1996ek,Mueller:1998fv,Radyushkin:1997ki}  by 
  \begin{align}
  \langle p' |  {\cal O}^\lambda (z) |p \rangle 
  = 2 {\cal P}^\lambda
  \int_{-1}^1 dx \, e^{-i x ({\cal P}z)} \, H(x,\xi,t;\mu^2) \ , 
\,  
 \
 \label{Hxi}
\end{align} 
where ${\cal O}^\lambda (z)= \bar   \psi(-z/2) \gamma^\lambda \hat W(-z/2,z/2) \psi (z/2)$ 
is the quark bilocal operator with $\hat W(-z/2,z/2; A) $ being Wilson line in the  fundamental
representation, 
 the coordinate  $z$ has only the $z_-$ light-cone component
 and \mbox{$\gamma^\lambda=\gamma^+$. } 
The matrix element is singular on the light cone, so one should 
use some regularization for it specified by  a scale $\mu$. 
For brevity, we will skip reference to $\mu^2$ in what follows. 
 
The invariant momentum transfer is given by \mbox{$t=(p-p')^2$.}
In principle, the r.h.s. of Eq. (\ref{Hxi}) 
has also the $r^\lambda$ term, where $r=p-p'$ is the momentum transfer. 
However, the GPD convention is to write $r^+=2\xi {\cal P}^+$,
where $\xi$ is the 
 skewness variable, and the two terms are combined in one GPD $H(x, \xi,t)$. 
 
 A similar definition  holds for  nucleons, 
  \begin{align}
 &  \langle p',s' | {\cal O}^+ (z)|p,s \rangle 
=
  \int_{-1}^1 dx \, e^{-i x {\cal P}^+z_-} \,  \nn &  \times  \left [  (\bar u' \gamma^+ u)  H(x,\xi,t)  - 
  \frac1{2M}  (\bar u' i \sigma^{+\mu } r_\mu u \, )E(x,\xi,t) \right ] \,  ,
 \
 \label{HxiN}
\end{align} 
where $\bar u' \equiv \bar u(p',s')$ and $u \equiv  u(p,s) $ are the nucleon spinors, while 
$H(x,\xi,t)$  and $  E (x,\xi,t) $  are the nucleon GPDs. 

One may  re-write these definitions in a more covariant  form that uses Lorentz invariants $({\cal P}z)$ and 
$(rz)$ only. For pion, we have 
  \begin{align}
&\left.  \langle p_2 | z_\lambda  {\cal O}^\lambda (z) |p_1 \rangle \right |_{z^2=0}
  \nn & 
  = 2( {\cal P} z)
  \int_{-1}^1 dx \, e^{-i x ({\cal P}z)} \, H(x,\xi,t;\mu^2) \Big | _{z^2=0}\  
\ . 
 \
 \label{HxiC}
\end{align} 

For nucleons, we have two GPDs
  \begin{align}
 & \left. \langle p',s' |  z_\lambda {\cal O}^\lambda (z)|p,s \rangle \right |_{z^2=0}
 \nn &
=
  \int_{-1}^1 dx \, e^{-i x ({\cal P}z)} \,   \Big \{  (\bar u' \slashed z u)   H(x,\xi,t) 
  \nn & 
 - \frac1{2M}  (\bar u' i \sigma^{z  r } u) E(x,\xi,t) \Big \} _{z^2=0} \,   .
 \
 \label{HxiNC}
\end{align}

\subsection{Double distribution description}

An alternative approach to describe nonforward matrix elements is based 
on DD formalism \cite{Mueller:1998fv,Radyushkin:1996nd,Radyushkin:1996ru,Radyushkin:1998es,Radyushkin:1998bz}. 
 Its guiding  idea  
 is to treat $P^+$ and $r^+$    as independent variables 
 and organize the plus-momentum flux as  a ``superposition'' 
of $P^+$ and $r^+$ momentum flows. 

\begin{figure}[t]
\centerline{\includegraphics[height=4cm]{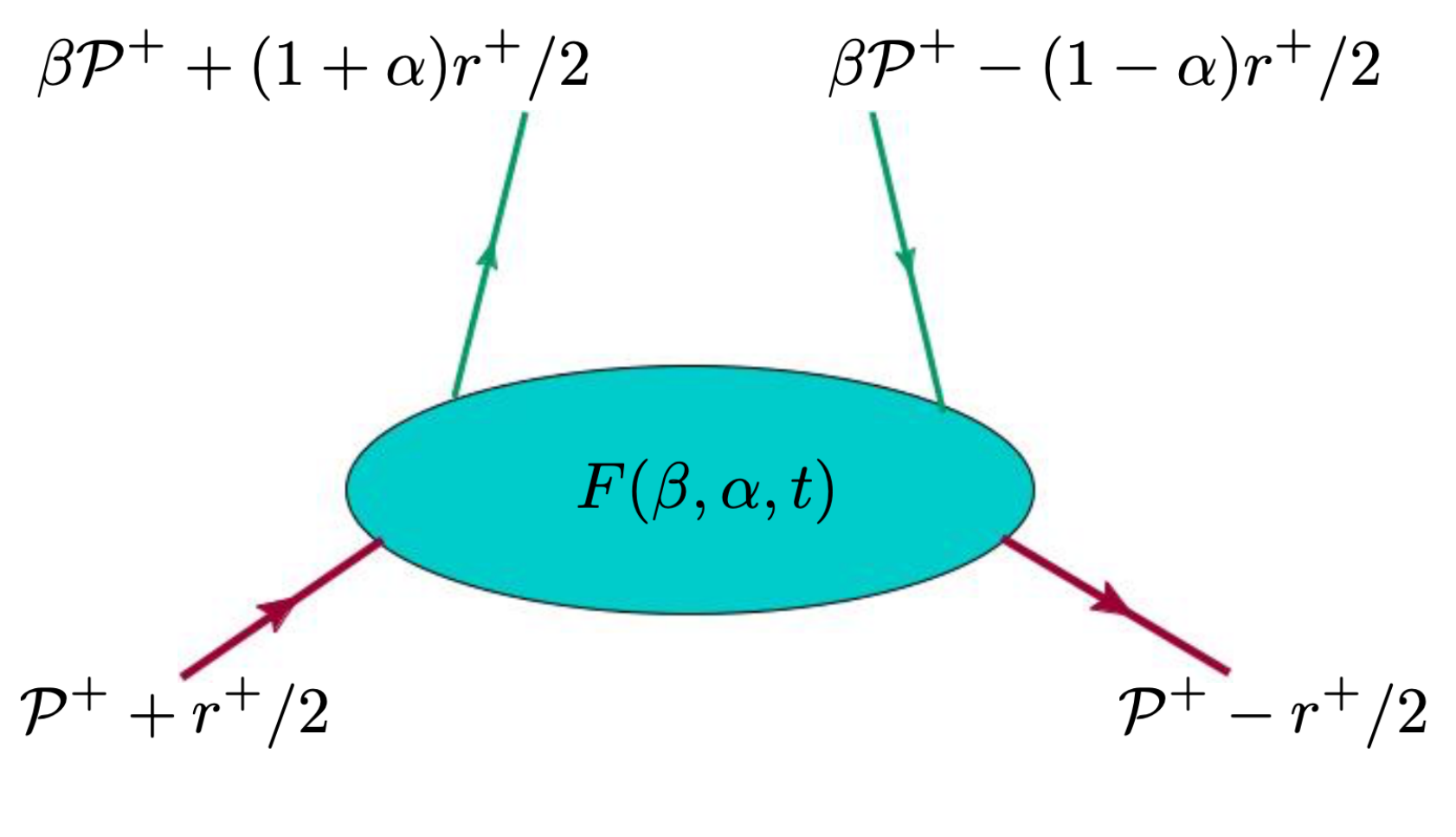}}
\caption{Flux of the momentum plus-components in terms of DD variables. }
\label{FluxDD}
\end{figure}

The parton 
momentum in this picture  is written as \mbox{$k^+ = \beta {\cal P}^+ + (1+\alpha) r^+/2$,}  i.e.,   
as a  sum of the  component $\beta {\cal P}^+$ due to 
the average hadron momentum $P$ (flowing in the $s$-channel)
and the  component $(1+\alpha) r^+/2$ due to the $t$-channel
momentum $r$, see Fig.~\ref{FluxDD}.

Thus, the $\alpha$-dependence of the DD $F(\beta,\alpha)$  
describes the distribution of the momentum transfer $r^+$
between  the initial and final quarks in fractions $(1+\alpha)/2$ and \mbox{$(1-\alpha)/2$.}
One may expect that it has a shape similar to those 
of parton   distribution {\it amplitudes}, e.g., with maximum at $\alpha=0$ (equal sharing of $r^+$)
and vanishing at  kinematical boundaries.  These  are located at \mbox{$\alpha =\pm (1-|\beta|)$}, 
since the support region for DDs is $|\alpha|+|\beta|\leq 1$ \cite{Radyushkin:1998bz}.

\subsubsection{Pion}

In terms of DDs, the matrix element (\ref{HxiC}) is written as \cite{Mueller:1998fv,Radyushkin:1996nd,Radyushkin:1998bz,Polyakov:1999gs} 
  \begin{align}
 &  \langle {\cal P} -r/2 | z_\lambda  {\cal O}^\lambda (z)|{\cal P} +r/2  \rangle _{z^2=0}
  \nn & 
 =
  \int_{\Omega} d\alpha d \beta  \,\, e^{-i \beta  ({\cal P}z) -i\alpha (rz)/2} \,   \nn &   \times  \Big \{  2 ({\cal P}z) F(\beta, \alpha,t)  
+ (rz) 
 \,  G(\beta, \alpha,t) \Big \} \Big | _{z^2=0} \ , 
\,  
 \
 \label{DD}
\end{align}  
where $\Omega$ is  the DD support region,  i.e.,   a rhombus in the $(\alpha\beta)$-plane
defined by $|\alpha|+|\beta| \leq 1$. 
The time reversal invariance requires that $F(\beta,\alpha,t)$ is  an even function of $\alpha$, 
while  $G(\beta,\alpha,t)$  is odd in $\alpha$.

Expanding $e^{-i \beta  ({\cal P}z) -i\alpha (rz)/2}$   in powers of  $({\cal P}z)$  and  $(rz)$, one observes 
that the  generic term  $({\cal P}z)^{N-k} (rz)^k$ may be obtained both  from $F$- and $G$-parts \cite{Teryaev:2001qm},
with two exceptions. Namely, one cannot obtain the $({\cal P}z)^{N}$ term from the $G$-part,
and one cannot obtain the $(rz)^N$  term from the $F$-part. 
The usual convention is to absorb all  the $({\cal P}z)^{N-k} (rz)^k$ terms with $k <N$  into the $F$-function, 
leaving  the $(rz)^N$  terms in the \mbox{$G$-function \cite{Polyakov:1999gs}.}  As a result, the $G$-part would not depend on  $({\cal P}z)$, 
and one can write 
  \begin{align}
 &  \langle {\cal P} - r/2 | z_\lambda  {\cal O}^\lambda (z)|{\cal P} +r/2  \rangle _{z^2=0}
  \nn & 
 =
 \Big \{     2 ({\cal P}z)   \int_{\Omega} d\alpha d \beta  \,\, e^{-i \beta  ({\cal P}z) -i\alpha (rz)/2} \,   F(\beta, \alpha,t)  
\nn &   
+ (rz)   \int_{-1} ^1 d\alpha  \, e^{-i\alpha (rz)/2}  
 \,  D (\ \alpha,t) \Big \} \Big | _{z^2=0} \ , 
\,  
 \
 \label{DDD}
\end{align}  
where $ D(\alpha,t) $ is 
 the $D$-term function  introduced in \mbox{Ref.  \cite{Polyakov:1999gs}.}
 It is odd in $\alpha$.

Comparing GPD and DD  parameterizations (\ref{HxiC}) and (\ref{DDD})   we get the relation between 
the pion GPD and DD
\cite{Mueller:1998fv,Radyushkin:1997ki,Polyakov:1999gs} 
  \begin{align}
    { H} \left (x,\xi,t\right )=&
     \int_{\Omega} d\alpha d \beta  \, 
   \delta (x- \beta-\alpha \xi)  F(\beta, \alpha,t)   \nn & 
   + {\rm sgn} (\xi )  D(x/\xi,t;\mu^2) \nn & 
   \equiv H_{DD}  +D
\,  \,
 \
 \label{HDD}
\end{align} 
As noticed in Ref. \cite{Teryaev:2001qm},  the $(\alpha\beta)$-integral above, i.e. the ``DD part''
$H_{DD}(x,\xi,t)$,  may be treated as the  Radon transform 
of  $F$. 

\subsubsection{Nucleon}

In the nucleon case, we have the following representation  \mbox{ \cite{Mueller:1998fv,Radyushkin:1996nd,Radyushkin:1998bz} } 
 \begin{align}
 &  \langle {\cal P} -r/2, s' | z_\lambda {\cal O}^\lambda (z))|{\cal P} +r/2 , s \rangle _{z^2=0} 
  \nn & 
 =
  \int_{\Omega} d\alpha d \beta  \,\, e^{-i \beta  ({\cal P}z) -i\alpha (rz)/2} \, 
    \nn & \times  \left [  ( \bar u' \slashed z  u) \,  h(\beta, \alpha,t) \ 
- \frac1{2M}  (\bar u' i \sigma^{z  r } u) e(\beta, \alpha,t) 
 \right ]\ 
  \nn & 
     + (rz)  \frac{(\bar u' u)}{M}  
  \int_{-1}^1 d\alpha  \,\, e^{ -i\alpha (rz)/2} \,  D(\alpha,t) \ . 
\,  
 \
 \label{DDN}
\end{align} 
Here,  $ h(\beta, \alpha,t)$ and $ e(\beta, \alpha,t)$ are even functions of $\alpha$, while $D(\alpha)$ is odd. 
Using 
  Gordon decomposition
\begin{align}
 	&
 \frac{{\cal P}^\lambda }{M}   \bar u'   u =  	 \frac1{2M}  \bar u' i \sigma^{\lambda r } u +  \bar u'  \gamma^\lambda  u \  , 
 \label{GD}
 	\end{align}
	and 
comparing (\ref{DDN})  with the GPD representation  (\ref{HxiNC}),  
gives  relation between the nucleon GPDs,  DDs and $D$-term \mbox{\cite{Goeke:2001tz}}.
Namely, we have 
    \begin{align}
    { H} \left (x,\xi,t\right )&=
     \int_{\Omega} d\alpha d \beta  \, 
   \delta (x- \beta-\alpha \xi)  h (\beta, \alpha,t) 
     \nn  & 
   + {\rm sgn} (\xi )  D(x/\xi,t) \nn
   &  \equiv H_{DD}  +D
\,  ,
 \
 \label{HNDD}
 \end{align} 
 for  $  { H} \left (x,\xi,t\right )$, and 
     \begin{align}
   { E} \left (x,\xi,t\right )&=
     \int_{\Omega} d\alpha d \beta  \, 
   \delta (x- \beta-\alpha \xi)  e (\beta, \alpha,t)   \nn & 
   - {\rm sgn} (\xi )  D(x/\xi,t) 
    \nn
   & \equiv E_{DD}  -D
\,  
 \
 \label{ENDD}
\end{align} 
 for  $  { E} \left (x,\xi,t\right )$.

Again, we may talk about  the ``DD parts'' $H_{DD}(x,\xi,t)$ and $E_{DD}(x,\xi,t)$ 
of the corresponding GPDs. Note that the $D$-term cancels in the sum  $H(x,\xi,t) + E(x,\xi,t) \equiv A(x,\xi,t)$.  
So, $A(x,\xi,t)$ is  built purely  from the DD  $a(\beta, \alpha,t) \equiv h(\beta, \alpha,t)+ e(\beta, \alpha,t)$.

\subsection{Fixed parity cases}

Usually we are interested in the functions corresponding to operators 
  \begin{align}
  {\cal O}^\lambda_{\pm}  (z) = \frac12 \left [  {\cal O}^\lambda (z)\pm  {\cal O}^\lambda (-z,A)\right ] \, 
    \label{OS}
\end{align} 
that are symmetric or antisymmetric with respect to the  inversion of $z$.  
These combinations  appear when we consider 
 ``nonsinglet''  $q-\bar q$ or ``singlet'' $q+\bar q$ parton distributions, respectively.  Since the $D$-term contribution
 (without  the overall $(rz)$ factor)
 is  odd in $z$, it  appears in the ``singlet'' case only.  However, the $H+E$ sum  does not contain 
 the $D$-term even in the singlet case.
 
 In fact, it is sufficient to consider matrix element of the original $ {\cal O}^\lambda (z)$ operator.
 The  real part of this matrix element is even in $z$ while its imaginary part is odd  in $z$.

\setcounter{equation}{0} 

\section{Some properties of GPDs and DDs} 

\subsection{DD-parts of GPDs} 

In this  section, we consider the relations between the DDs and the ``DD parts'' of GPDs which they generate, thus 
ignoring for a while the 
 $D$-term contributions to GPDs. 
The   $D$-term will be  discussed later in  the paper.
For definiteness, we will have in mind    relations between the DD part of the  pion GPD and its DD.
All the relations  are equally applicable to  the DD parts of the  nucleon GPDs. 

\begin{figure}[h]
\centerline{\includegraphics[height=5.5cm]{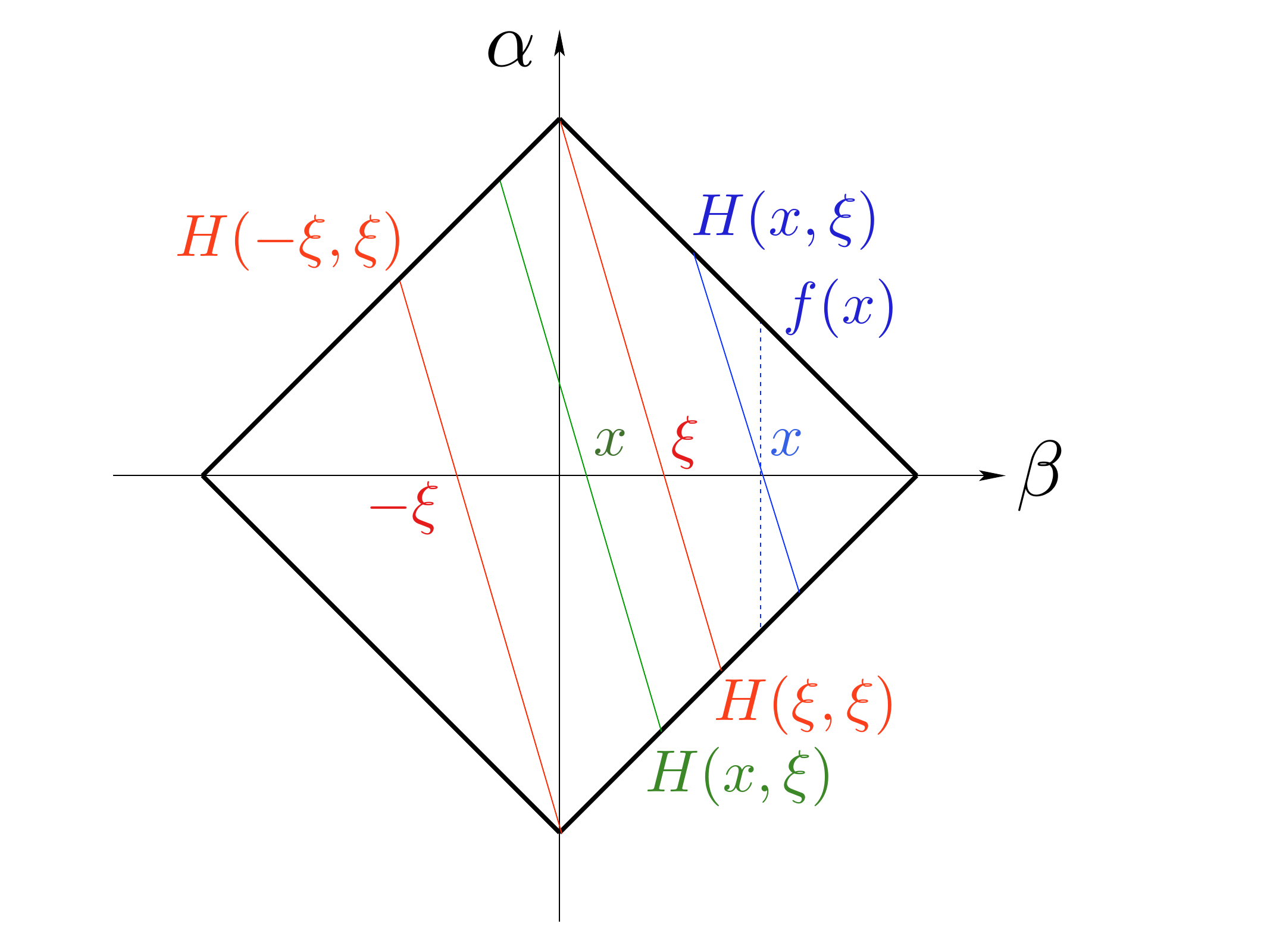} 
 }
\caption{DD support rhombus and  integration lines  producing 
the DD parts of $H(\xi,\xi) $, 
$H(-\xi,\xi) $, $H(x,\xi=0)=f(x)$ and  $H(x,\xi)$ in  ($|x|>\xi$) and    
 ($|x|<\xi$)  regions.
   }
\end{figure}

\subsection{$\xi=0$ limit}

Taking $\xi=0$, we have 
  \begin{align}
    {H}  \left (x,\xi=0,t\right )& =
   \int_{-1} ^1   d \beta  \,   \delta (x- \beta) 
      \int_{-1+|\beta|} ^{1-|\beta|}  d\alpha
  F(\beta, \alpha,t)
  \nn & 
  =     \int_{-1+|x|} ^{1-|x|}  d\alpha
  F(x, \alpha,t) \equiv f(x,t) 
\,  \, . 
 \
 \label{Htxi}
\end{align} 
This means that integrating $ F(\beta, \alpha,t)$ 
over vertical lines $\beta=x$ gives the $\xi=0$ (``non-skewed'') GPD $ {\cal H} \left (x,\xi=0,t\right )$,
which we will also denote as $f \left (x,t\right )$. 
It is the simplest GPD, that was called  ``nonforward parton density''  in the paper  \cite{Radyushkin:1998rt},
where it has been introduced. 
  It differs from the forward PDF $f(x)$ 
by the presence of  $t$-dependence and satisfies $f \left (x,t=0\right )=f(x)$.

\subsection{Polynomiality}

The DD representation automatically produces a GPD satisfying 
the polynomiality property. 
Indeed, 
 \begin{align}
  & 
  \int_{-1}^1 dx \,  x^n    {H}_{DD} \left (x,\xi,t\right )
  \nn & =
   \int_{-1}^1 dx  \, x^n      \int_{\Omega} d\alpha d \beta  \,
   \delta (x- \beta-\alpha \xi)   F(\beta, \alpha,t)
   \nn & =
     \sum_{k=0}^n \frac{n!}{k! (n-k)!}  \,  \xi^k  
     \int_{\Omega} d\alpha d \beta  \,
 \beta^{n-k} \alpha^k   F(\beta, \alpha,t) 
\,  ,
 \
 \label{Htxin}
\end{align} 
i.e., the $n^{\rm th}$ $x$-moment of  $ {H}_{DD} \left (x,\xi,t\right )$
is a polynomial in $\xi$ of the order not exceeding $n$. 

Note that, since $F(\beta, \alpha,t) $ is even in $\alpha$, the summation over $k$
involves even $k$ only, i.e. (\ref{Htxi}) is in fact an expansion in powers of $\xi^2$. 

\subsection{Ioffe-time distributions} 

By Lorentz invariance, the 
matrix element  (\ref{HxiC}) defining GPD is 
a function of the scalars $(pz)\equiv -\nu_1 $ and $(p'z)\equiv -\nu_2$,
two Ioffe-time parameters, so we may write 
  \begin{align}
  \langle p_2 | \bar   \psi(-z/2) \slashed z 
    \psi (z/2)|p_1 \rangle = 2( {\cal P}z)  {I} (\nu_1,\nu_2,t)
\,   ,
 \
 \label{Mnn}
\end{align} 
where ${I} (\nu_1,\nu_2,t)$ is the double Ioffe-time distribution. 
Since $z=z_-$ is assumed, 
 only  the values of the plus components of   momenta
are  essential in the scalar products $(p_1z)$ and $(p_2z)$.   
The skewness variable  $\xi$ in this case is given by 
  \begin{align}
  \xi 
  =\frac{\nu_1-\nu_2}{\nu_1+\nu_2} \equiv \frac{\nu_1-\nu_2}{2\nu}
\,  .
 \
 \label{Zz1z2}
\end{align} 
We have introduced here the variable $\nu =(\nu_1+\nu_2)/2$. 
Treating $\nu$ and $\xi$ as independent variables, we 
define the  {\it generalized Ioffe-time distribution} (GITD)  as
\begin{align}
  &{I} (\nu_1,\nu_2,t) = {\cal I} (\nu, \xi ,t)\ . 
  \end{align} 
According to (\ref{Hxi}), it is a Fourier transform of the   GPD
  \begin{align}
  &  {\cal I} (\nu, \xi,t)
 =  \int_{-1}^1 dx \, e^{i x \nu } \, 
    {H} \left (x,\xi,t\right )
\,  . 
 \
 \label{MH}
\end{align} 
Using Eq. (\ref{DD}), we can write the DD part of  GITD in terms of DD
  \begin{align}
  &  {\cal I}_{DD}  (\nu, \xi,t)
 =
   \int_{-1} ^1   d \beta  \,
  e^{i  \nu \beta } \,      \int_{-1+|\beta|} ^{1-|\beta|}  d\alpha \,  e^{i  \nu \alpha \xi }   F(\beta, \alpha,t)
 \ . 
 \label{MF}
\end{align}

\subsection{DD profile and $\xi$-dependence}

If $ F(\beta, \alpha,t)$ has an infinitely narrow profile 
in the \mbox{$\alpha$-direction,}  i.e. if $ F(\beta, \alpha,t) = f(\beta,t) \delta (\alpha)$,
then the  \mbox{$\xi$-dependence}  disappears, and we deal with the simplest GPD
$f(x,t)$. 
A nontrivial dependence on the skewness $\xi$ is  obtained if  DD has a nonzero-width  profile 
in the $\alpha$-direction. 

Using the DD representation (\ref{MF})  for the GITD  and 
expanding $e^{i  \nu \alpha \xi } $ into Taylor series,   we  get the following   expansion in powers of $\xi^2$ 
  \begin{align}
  &  {\cal I}_{DD}  (\nu, \xi,t)
 = 
   \int_{-1} ^1   d \beta  \,
  e^{i  \nu \beta } \,      \int_{-1+|\beta|} ^{1-|\beta|}  d\alpha  F(\beta, \alpha,t)
   \nn & - \frac{\xi ^2 \nu^2 }{2}
   \int_{-1} ^1   d \beta  \,
  e^{i  \nu \beta } \,      \int_{-1+|\beta|} ^{1-|\beta|}  d\alpha \,   \alpha^2    F(\beta, \alpha,t) +\ldots 
 \label{MF3}
\end{align} 
(odd powers of $\xi$ do not appear because $F(\beta, \alpha,t)$ is even in $\alpha$).
By analogy with  (\ref{Htxi}), we will use notation   $ f_2(\beta, t)$
for  the second $\alpha$-moment of   $ F(\beta, \alpha,t) $ 
  \begin{align}
  & 
   \int_{-1+|\beta|} ^{1-|\beta|}  d\alpha \,   \alpha^2 \, F(\beta, \alpha,t) \equiv   f_2(\beta, t) \, 
 \label{2nd}
\end{align} 
As a result,  we write 
  \begin{align}
     {\cal I}_{DD} (\nu, \xi,t) 
 &= 
   \int_{-1} ^1   d \beta  \,
  e^{i  \nu \beta } \,    \Bigg \{  f(\beta, t)
   - \frac{\xi ^2 \nu^2 }{2}
   f_2(\beta, t) \Bigg \} 
    \nn & 
  + {\cal O} (\xi^4) 
   \nn & 
 \equiv 
 {\cal I}_0 (\nu, t) 
   - \frac{\xi ^2 \nu^2 }{2}
    {\cal I}_2(\nu, t) 
  + {\cal O} (\xi^4 )  \ .
 \label{MF4}
 \end{align} 

\section{Pseudodistributions} 

\subsection{Definitions}

On the lattice,  we 
choose  $z=z_3$, and 
  introduce pseudo-GPDs $ {\cal  H} \left (x,\xi,t;z_3^2\right )$
(and also $ {\cal  E} \left (x,\xi,t;z_3^2\right )$ in the nucleon case).

The two Ioffe-time parameters are given now by \mbox{$\nu_1=p^{(3)} z_3 \equiv P_1z_3$} 
and  $\nu_1=p'^{(3)} z_3 \equiv P_2z_3$. In terms of momenta $P_{1,2}$, 
the skewness $\xi$ is  given by 
\begin{align}
\xi= \frac{P_1-P_2}{P_1+P_2} \ . 
\end{align} 
The pseudo-GITD will be denoted as ${\cal M} (\nu, \xi,t;z_3^2)$,
e.g.,  the 
 inverse transformation for $ {\cal  H}$  is written as 
\begin{align}
 {\cal  H} \left (x,\xi,t;z_3^2\right ) =&\frac1{2\pi} \int_{-\infty}^\infty d\nu \, 
 e^{-i x \nu } \, {\cal M} (\nu, \xi,t;z_3^2)
   \,     
\,  .
 \
 \label{HM}
\end{align} 
Similarly, to denote  pseudo-DDs, we will just add $z_3^2$ to their arguments. 

 \subsection{Contaminations} 

On the lattice, we have $z^2\neq 0$, and we need to  add extra $z$-dependent structures to the original
twist-2  parameterization 
 \begin{align}
 &z_\lambda M^\lambda\equiv     \langle {\cal P} -r/2, s' |  z_\lambda {\cal O}^\lambda (z)|{\cal P} +r/2 , s \rangle 
  \nn & 
 =
  \int_{\Omega} d\alpha d \beta  \,\, e^{-i \beta  ({\cal P}z) -i\alpha (rz)/2} \, 
    \nn & \times  \Bigg \{  ( \bar u' \slashed z  u) \, h(\beta, \alpha,t)  - \frac1{2M}  (\bar u' i \sigma^{z  r } u) e(\beta, \alpha,t) 
\nn & \hspace{4cm} +   \frac{ \bar u'   u }{M} (rz)  \delta (\beta) D(\alpha,t)  \Bigg \}\ 
  \nn & \equiv   ( \bar u' \slashed  z u)  H_{DD}   - \frac1{2M}  (\bar u' i \sigma^{z  r } u)   E_{DD} +(rz)  \frac{ \bar u'   u }{M}    D
  \ , 
\,  
 \
 \label{DDN3a}
\end{align} 
where $z^2=0$ and  the ITDs $H_{DD} , E_{DD}  $ and $D$ are functions of $\nu$, $\xi$  and $t$.

A classification of  additional  tensor structures that appear in  parametrizations of  $M^\lambda$
off the light cone was done in Ref. \cite{Bhattacharya:2022aob}, where 
eight  independent structures have been identified.

However, there is some subtlety  related to the fact that,    for lattice extractions, we need to  parametrize   the ``non-contracted'' matrix element $M^\lambda$.
In this case, the index $\lambda$ in local operators $\bar \psi \gamma^\lambda (zD)^N \psi$ 
is not symmetrized with the indices $D^{\mu_1}\ldots D^{\mu_N}$ in covariant derivatives,
which is necessary for building twist-2 local operators.

A  way to perform symmetrization for  bilocal operators was indicated in Ref. \cite{Balitsky:1987bk}.
Further studies of 
 parameterizations for matrix elements with an open index have been done  in 
 \mbox{Refs. \cite{Anikin:2000em,Penttinen:2000dg,Belitsky:2000vx,Radyushkin:2000jy,Radyushkin:2000ap,Kivel:2000cn}. }
An important observation made there is that $M^\lambda$ {\it should } contain terms that vanish when contracted 
with $z_\lambda$, such as $r^\lambda ({\cal P}z) - {\cal P}^\lambda (rz)$.
One can see that \mbox{$r^\lambda - {\cal P}^\lambda (rz)/ ({\cal P}z)\equiv \Delta_\perp^\lambda$} 
 is the part of the momentum transfer $r$ 
that is transverse to $z$. 

As shown in these papers, one should add  Wandzura-Wilczek-type (WW)  terms \cite{Wandzura:1977qf}
to  parametrizations of GPDs  to secure electromagnetic  gauge invariance of the DVCS amplitude
\cite{Guichon:1998xv}  with
${\cal O}(\Delta_\perp)$ accuracy.  
While the WW terms are ``kinematical twist-3'' contributions built from twist-2 GPDs,
one cannot exclude non-perturbative (dynamical) twist-3 terms accompanied by the $\Delta_\perp^\lambda$
factor. 

 Among   additional structures listed in Ref. \cite{Bhattacharya:2022aob}, 
one can see the structure $(\bar u' i \sigma^{\lambda z} u )$ that also vanishes when multiplied by $z_\lambda$, 
and thus should be treated as a ``higher-twist'' term. 

On the other hand,  two other additional structures, 
$(\bar u' i \sigma^{ z r}  {\cal P}^\lambda u)$ and $(\bar u' i \sigma^{ z r}  {r}^\lambda u)$, after contraction   
with $z_\lambda$,  
produce the  same ``twist-2'' structure \mbox{$\sim \sigma^{ z r} $} that accompanies the $E_{DD}$ contribution.
In this sense, the invariant amplitudes accompanying these structures, have a \mbox{twist-2 }  component.

Note, however, that 
 combinations \mbox{$\sigma^{ z r}  {\cal P}^\lambda - \sigma^{ \lambda r}  ({\cal P}z) $}  
and $\sigma^{ z r}  {r}^\lambda - \sigma^{ \lambda r}  (rz) $   vanish after  contraction with   $z_\lambda$.
So, we propose to  use these ``subtracted'' forms in building    the basis of additional terms, rather than 
just $\sigma^{ z r}  {\cal P}^\lambda $ and $\sigma^{ z r}  {r}^\lambda $. 
Since the ``subtracted''   structures do not contribute to  the \mbox{twist-2}  parameterization (\ref{DDN3a}), 
  the DDs associated with them  should   be  classified as ``higher-twist'' ones. 
  
For this reason,  we construct 
a   parameterization for  $M^\lambda$ in which \mbox{``twist-2''}  and ``higher-twist'' terms are explicitly separated,
 \begin{align}
 &M^\lambda=   ( \bar u' \gamma^\lambda u)   H_{DD} 
 - \frac1{2M}  (\bar u' i \sigma^{\lambda  r } u)   E_{DD}  %
 + r^\lambda \frac{ \bar u'   u }{M} D 
 \nn & +  \left [r^\lambda { ({\cal P}z)} - {\cal P}^\lambda {(rz)}\right ] \frac{ \bar u'   u }{M}  Y
\nn  &-  
\frac{1}{M} \left [(\bar u' i \sigma^{ z r} u ){\cal P}^\lambda  - (\bar u' i \sigma^{ \lambda r} u )({\cal P}z) \right ]  X_1
 \nn  &-  
\frac{1}{M} \left [(\bar u' i \sigma^{ z r} u ){r}^\lambda  - (\bar u' i \sigma^{ \lambda r} u )({r}z) \right ]  X_2
\nn &+ { (\bar u' i \sigma^{\lambda z} u )}{M} X_3
\nn  &  +i({ \bar u'   u }){M} z^\lambda Z_1-  {(\bar u' i \sigma^{ z r} u )}  Mz^\lambda Z_2
  \  .
\,  
 \
 \label{DDN3n}
\end{align}
We have here  $Y$ and $X_i$ terms whose contribution vanishes 
when contracted with $z_\lambda$, and 
 $Z_i$ terms that produce $z^2$ factor after contraction with $z_\lambda$.
 
Formally, the $r^\lambda { ({\cal P}z)} - {\cal P}^\lambda {(rz)}$  combination  does not contain new   structures that are independent 
from those present in the ``twist-2'' line.
However, the corresponding invariant amplitude, which is denoted as $Y$, is generated by a new DD. This 
``higher-twist'' DD
 is different from the twist-2 DDs $h(\beta,\alpha)$, $e(\beta,\alpha)$ and the $D$-term, which are also 
associated with  the $\sim \bar u'   u$ structures. 

Of course, 
using Gordon decomposition 
\begin{align}
 	&
 \frac{{\cal P}^\lambda }{M}   \bar u'   u =  	 \frac1{2M}  \bar u' i \sigma^{\lambda r } u +  \bar u'  \gamma^\lambda  u \  , 
 \label{GD2} 
 	\end{align}
	 we can re-write (\ref{DDN3n}) in a form explicitly having just eight structures
 \begin{align}
 &M^\lambda=   ( \bar u' \gamma^\lambda u)   \left [ H_{DD}  - (rz) Y\right ] + r^\lambda \frac{ \bar u'   u }{M}[ D   + ({\cal P}z)Y] \nn & 
 - \frac1{2M}  (\bar u' i \sigma^{\lambda  r } u)   [E_{DD}   +(rz) Y] 
  \nn  & 
-  
\frac{1}{M} \left [(\bar u' i \sigma^{ z r} u ){\cal P}^\lambda  - (\bar u' i \sigma^{ \lambda r} u )({\cal P}z) \right ]  X_1
 \nn  &-  
\frac{1}{M} \left [(\bar u' i \sigma^{ z r} u ){r}^\lambda  - (\bar u' i \sigma^{ \lambda r} u )({r}z) \right ]  X_2
\nn &+ { (\bar u' i \sigma^{\lambda z} u )}{M} X_3
\nn  &  +i({ \bar u'   u }){M} z^\lambda Z_1-  {(\bar u' i \sigma^{ z r} u )} Mz^\lambda  Z_2
  \ , 
\,  
 \
 \label{DDN3c}
\end{align}
like in Ref. \cite{Bhattacharya:2022aob}.  
 To establish a direct  correspondence, we note that   
 Ref. \cite{Bhattacharya:2022aob} uses a  basis  in    which $ ( \bar u' \gamma^\lambda u) $ is substituted by
two other structures that appear in the Gordon decomposition (\ref{GD2}). 
Also, all the terms containing $ (\bar u' i \sigma^{\lambda  r } u) $ are combined in one contribution.
Using this basis, we have
 \begin{align}
 &M^\lambda=   \frac{{\cal P}^\lambda }{M}  ( \bar u'   u )
 [H_{DD}  -(rz) Y] + r^\lambda \frac{ \bar u'   u }{M}[ D   + ({\cal P}z)Y] \nn & 
 - \frac1{2M}  (\bar u' i \sigma^{\lambda  r } u)   [H_{DD} + E_{DD} +  2({\cal P}z)X_1 + 2 (rz) X_2 ] 
  \nn  & 
+ { (\bar u' i \sigma^{\lambda z} u )}{M} X_3  +i ({ \bar u'   u }){M}  z^\lambda Z_1 \nn  &-  \frac{(\bar u' i \sigma^{ z r} u )}{M} [{\cal P}^\lambda   X_1
+ r^\lambda  X_2+ z^\lambda M^2 Z_2]
  \  .
\,  
 \
 \label{DDN3d}
\end{align}
Comparing Eq. (\ref{DDN3d})  with the coefficients $A_i$  in \mbox{Eq. (35) } of Ref. \cite{Bhattacharya:2022aob}, we 
establish the correspondence \mbox{ $A_1=  [H_{DD}  -(rz) Y] $,} $A_2=iZ_1$, $-A_3 = D   + ({\cal P}z)Y$,
$A_4=-X_3$, $A_5 = (H_{DD}+E_{DD})/2+({\cal P}z)X_1 +  (rz) X_2$, $A_6=X_1$, $A_7= Z_2$, $A_8 = -X_2$. 

The main difference is that $H_{DD}$ and $D$  contributions  in   Eq. (\ref{DDN3d})   come with the contamination from the \mbox{$Y$-function,}
the $9{\rm th}$ ``higher-twist''  ITD. 
Also,  the $ (\bar u' i \sigma^{\lambda  r } u) $ structure is accompanied by a factor in which the $Y$ term is absent,
but there are contaminations from $X_1$ and $X_2$  contributions.

\setcounter{equation}{0} 
 
 \section{Fitting  pseudodistributions}

   \subsection{Nonforward parton pseudo-density  $f(\beta,t,z_3^2)$}
   
   Taking $\xi=0$ we have
    \begin{align}
  &   {\cal M} (\nu, \xi=0,t;z_3^2)=    \int_{-1} ^1 d \beta  \, e^{i  \nu \beta } \,    f(\beta,t,z_3^2 )  \ ,
  \label{xizero0}
      \end{align} 
      where $\nu =P_1z_3=P_2 z_3$.  An important point is that $\xi=0$ may be achieved for different pairs of equal 
   initial and final momenta $P_1=P_2\equiv P$. One should check that lattice gives the same 
   curve for  different $P$'s,  up to evolution-type dependence on $z_3^2$.
   
   One can use  relation (\ref{xizero0}) to fit $ f(\beta,t,z_3^2)  $. First, taking $t=0$, we fit the forward  pseudodistribution $f(\beta,z_3^2)$,
   just as a pseudo-PDF. After that, one can  vary $t$, by changing the transverse components 
   $\Delta_{\perp}^{1,2 }$,   for several fixed $\nu$.
   In this way, one  can study what kind of dependence on $t$ we have (dipole, monopole, etc.), and how it changes with $\nu$.

   \subsection{$\alpha^2$-moment function   $f_2(\beta,t,z_3^2)$}
   
   The next step is to check if  the $\xi$-dependence 
   of the lattice data for $ {\cal M} (\nu, \xi,t;z_3^2)$  agrees with  the form 
    \begin{align}
  & {\cal M} (\nu, \xi,t;z_3^2)=   {\cal M} (\nu, \xi=0,t;z_3^2) 
  \nn & 
    - \frac{\xi ^2 \nu^2 }{2}  {\cal M}_2 (\nu, t;z_3^2) +   {\cal O} (\xi^4) \ ,
  \label{Mexp}
      \end{align} 
and extract $f_2(\beta,t,z_3^2)$  using 
  \begin{align}
    {\cal M}_2 (\nu, \xi,t;z_3^2) 
 &= 
   \int_{-1} ^1   d \beta  \,
  e^{i  \nu \beta } \,     
   f_2(\beta, t;z_3^2) 
   \ .
 \label{MF42}
 \end{align} 
The $\alpha$-dependence of the DD $F(\beta,\alpha)$  
describes the distribution of the momentum transfer $r=P_1-P_2$
between  the initial and final quarks. 
It is expected that it has a shape similar to those 
of parton distribution amplitudes.

\subsection{Factorized DD Ansatz} 

A  nonzero-width  profile 
of DD in the $\alpha$-direction may be modeled   by using the 
Factorized DD Ansatz \cite{Radyushkin:1998es,Radyushkin:1998bz}
\begin{align}
F_N(\beta,\alpha,t) =f(\beta,t)  \frac{\Gamma
   \left(N+\frac{3}{2}\right)}{\sqrt
   {\pi } \Gamma (N+1)}
\frac{[(1-|\beta|)^2 -\alpha^2]^{N} \, 
 }{ 
(1-|\beta|)^{2N +1} } \  ,
\label{eq_profile} 
\end{align}
with $N$  being the parameter governing the width of the $\alpha$-profile of the model DD  $F_N(\beta,\alpha,t) $.
The $\alpha$-integral of $F_N(\beta,\alpha,t)  $ gives the nonforward parton density $f(\beta,t) $. 

The $[(1-|\beta|)^2 -\alpha^2]$ factor reflects the support properties 
of the DD, which vanishes if $|\beta|+|\alpha| >1$. 
The Ansatz also complies with the requirement  that $F(\beta,\alpha)$ should be an even function of $\alpha$. 

For $f(\beta,t)$ one can also take a factorized  form  $f(\beta,t)=f(\beta) F(t)$,
where $f(\beta) $ is the forward PDF, and $F(t)$ some form factor.
Combining (\ref{MF}) and (\ref{eq_profile}) gives 
  \begin{align}
  & {\cal M}_N (\nu, \xi,t;z_3^2)=    \int_{-1} ^1 d \beta  \, e^{i  \nu \beta } \,    f(\beta,t;z_3^2)  
    \nn &\times 
     \int_{-1+|\beta|}^{1-|\beta|}  d\alpha  \,
  e^{i  \nu \alpha \xi  } \,    \frac{\Gamma
   \left(N+\frac{3}{2}\right)}{\sqrt
   {\pi } \Gamma (N+1)}
\frac{[(1-|\beta|)^2 -\alpha^2]^{N} \, 
 }{  
(1-|\beta|)^{2N +1} }  \ .
 \
 \label{Mtxi}
\end{align} 

Integral over $\alpha$ can be taken 
\begin{align}
&A_N(\beta) =  \int_{-1+|\beta|}^{1-|\beta|}  d\alpha  \,
  e^{i  \nu \alpha \xi  } \,     
\frac{[(1-|\beta|)^2 -\alpha^2]^{N} \, 
 }{ 
(1-|\beta|)^{2N +1} }\nn 
&=   \int_{-1}^{1}  d\eta  \,
  e^{i  \nu \xi  (1-|\beta|)\eta } \,    
(1-\eta^2)^{N} \, 
\nn &  =
  \,
   _0\tilde{F}_1\left(;N+\frac{3}{2} ;-\frac{\nu^2 \xi^2  (1-|\beta|)^2}{4}\right)
    \sqrt{\pi }  \Gamma(N+1) \ , 
\label{AN} 
\end{align}
where 
  \begin{align}
  &  
   _0\tilde{F}_1\left(; b;z\right)=
   \sum_{k=0}^\infty \frac{z^k}{
  \Gamma(b+k)k!}
   \end{align} 
   is a hypergeometric function. 

So, we have a model  for pseudo-GITD
  \begin{align}
  &  {\cal M}_N (\nu, \xi,t;z_3^2)=    \int_{-1} ^1 d \beta  \, e^{i  \nu \beta } \,    f(\beta,t)  
    \nn &\times 
      \,
   _0\tilde{F}_1\left(;N+\frac{3}{2} ;-\frac{\nu^2 \xi^2  (1-|\beta|)^2}{4}\right)
 {\Gamma
   \left(N+\frac{3}{2}\right)} \ ,
      \end{align} 
    or, expanding in $\xi$,
     \begin{align}
  &  {\cal M}_N (\nu, \xi,t;z_3^2)=    \int_{-1} ^1 d \beta  \, e^{i  \nu \beta } \,    f(\beta,t)  
    \nn &\times 
     \sum_{k=0}^\infty   \left ( -\frac{\nu^2 \xi^2  (1-|\beta|)^2}{4}\right)^k 
      \frac{\Gamma
   \left(N+\frac{3}{2}\right)}{k! \Gamma(N+3/2+k)} \ .
   \label{pGITD}
      \end{align} 
   This expansion may be  also obtained   by taking Taylor series of  $ e^{i  \nu \xi  (1-|\beta|)\eta } $ in Eq. (\ref{AN}),
   and integrating over $\eta$.

  \subsection{Check of polynomiality}
 
   Getting GPDs from a DD representation  guarantees that the resulting 
   GPD  has the polynomiality property. Still, we can double-check this. 
   Note that the $x^n$ moment ${\cal M}^{(n)} \left (\xi,t;z_3^2\right ) $  of a  pseudo-GPD $    {\cal H} \left (x,\xi,t;z_3^2\right )$ 
   is proportional to the coefficient accompanying $\nu^n$ in the 
Taylor expansion
    \begin{align}
  & {\cal M} (\nu, \xi,t;z_3^2)=   \sum_{N=0}^\infty \frac{i^n \nu^n}{n!}   {\cal M}^{(n)}  ( \xi,t;z_3^2)\ .
      \end{align} 
Now, from  
      \begin{align}
  &  {\cal M}_N (\nu, \xi,t;z_3^2)=    \int_{-1} ^1 d \beta  \, \sum_{m=0}^\infty
 \frac{ (i  \nu \beta)^m}{m!}  \,    f(\beta,t;z_3^2)  
    \nn &\times 
     \sum_{k=0}^\infty   \left ( -\frac{\nu^2 \xi^2  (1-|\beta|)^2}{4}\right)^k 
      \frac{\Gamma
   \left(N+\frac{3}{2}\right)}{k! \Gamma(N+3/2+k)}
      \end{align} 
      we see  that ``$n$'' in $ {\cal M}_N^{(n)}  $  corresponds here to $n=m+2k$. On the other hand,
      $ {\cal M}_N^{(n)}  ( \xi,t;z_3^2)  $
       is a polynomial in $\xi$ of order $2k$ which is equal or smaller than $n$ since $m\geq 0$.


  \subsection{Fitting $\alpha$-profile width}
  
  After fixing  $f(\beta,t;z_3^2)$ that gives 
   the profile of DD in the $\beta$-direction, we may quantify  what kind of profile it has in the $\alpha$-direction.
  The presence of  a nontrivial profile is indicated by the presence of $\xi$-dependence. 
   Using  the  first terms of the series for $ _0\tilde{F}_1\left(; b;z\right)$ 
       \begin{align}
  &  
 \Gamma(b)    _0\tilde{F}_1\left(; b;z\right)=
   \sum_{k=0}^\infty \frac{z^k  \Gamma(b)}{
  \Gamma(b+k)k!}=1+\frac{z}{b} +\frac{z^2}{2b(b+1)} + \ldots
   \end{align} 
   we write (\ref{pGITD}) as 
     \begin{align}
  &   {\cal M} (\nu, \xi,t;z_3^2;N)=    \int_{-1} ^1 d \beta  \, e^{i  \nu \beta } \,    f(\beta,t)  
    \Bigg  \{ 1-
      \frac{\nu^2 \xi^2  (1-|\beta|)^2}{4(N+3/2)}
      \nn & +  \left ( \frac{\nu^2 \xi^2  (1-|\beta|)^2}{4}\right)^2
      \frac{1}{
2(N+3/2)(N+5/2)}+\ldots \Bigg \} \ . 
\label{DDexp}
      \end{align}

  In Eq. (\ref{DDexp}), $\xi$ appears through the combination \mbox{$\xi \nu =(\nu_1-\nu_2)/2$.} On the lattice, we have
  $\nu_1=P_1 z_3$,   $\nu_2=P_2 z_3$. Hence,  the presence of a nontrivial 
  profile should be reflected by the dependence
  of the data on the difference $P_1-P_2$  for a fixed sum $P_1+P_2$.
  The first correction in Eq. (\ref{DDexp})  is given by 
    \begin{align}
   \delta  {\cal M} (\nu, \xi,t;z_3^2;N)= &  -
    \int_{-1} ^1 d \beta  \, e^{i  \nu \beta } \,    f(\beta,t;z_3^2)  (1-|\beta|)^2\nn &
    \times 
      \frac{\xi^2 \nu^2  }{4(N+3/2)} \ . 
\label{DDexp10}
      \end{align} 
  
  Using this expression, one  can try to   determine the profile parameter $N$.
  This task  probably will not be easy, since the correction looks rather small 
  due to a small overall   factor $\sim \xi^2/4$. 
  
  We may also estimate the extra suppression 
  due to the $ (1-|\beta|)^2$ factor in the integrand of (\ref{DDexp10}).
  For a simple illustration, take $f(\beta,t)=4(1-|\beta|)^3 $.
  In this case,
      \begin{align}
   \int_{-1} ^1 d \beta  \, e^{i  \nu \beta } \,    f(\beta,t) =&\frac{48}{\nu ^4}  \left( \cos (\nu
   )-1+\frac{\nu ^2}{2}\right)\nn &
   =2-\frac{\nu ^2}{15}+\frac{\nu
   ^4}{840}+O\left(\nu ^5\right)
      \end{align} 
while 
        \begin{align}
    \int_{-1} ^1 d \beta & \, e^{i  \nu \beta } \,    f(\beta,t)  (1-|\beta|)^2 
    \nn & 
= \frac{960}{\nu ^6}  \left(-\cos
   (\nu )+1-\frac{\nu ^2}{2} +\frac{\nu
   ^4}{24}\right)\nn &
   =\frac{4}{3}-\frac{\nu
   ^2}{42}+\frac{\nu
   ^4}{3780}+O\left(\nu ^5\right) \ .
      \end{align} 
      Thus, the additional suppression is by about $2/3$ for small $\nu$, i.e., not very strong. 
      
      \subsection{$D$-term} 
      
      When we take the $z$-odd  part $ {\cal O}_-^\lambda $ of the  operator ${\cal O}^\lambda (z)$,  its parametrization contains a nonzero 
      $D$-term. In GPD description, it appears in a mixture with $H_{DD}$ (and also $E_{DD}$ in the nucleon case) GPDs.  However, 
      using all possible helicity states for nucleons and various values of $\lambda$, one can 
      construct sufficient number of linearly independent relations.  To  
    separate the DDs  that appear in the parametrization of Eq. (\ref{DDN3c}) one can use, e.g., singular value decomposition technique. 
     Unfortunately, as seen from Eq. (\ref{DDN3c}), the $D$-term obtained in this way comes 
     together with the $Y$-contamination.

      Another  way is to eliminate $H_{DD}$, $E_{DD}$, etc.  contributions   from the matrix element of 
      $ {\cal O}_-^\lambda $ 
by 
   taking  kinematics in which 
      $({\cal P} z)=0$. As a result, $\alpha$-even DD $h(\beta, \alpha)$  will be integrated with  the $\alpha$-odd 
      function $\sin (\alpha (rz))$, etc.,
       so that  we will have 
        \begin{align}
 &  \langle {\cal P} -r/2, s' |  {\cal O}_-^\lambda (z)|{\cal P} +r/2 , s \rangle  |_{({\cal P} z)=0}
  \nn & 
 =
      r^\lambda \frac{(\bar u' u)}{M}  
  \int_{-1}^1 d\alpha  \,\, e^{ -i\alpha (rz)/2} \,  D(\alpha,t) 
  \nn &  + ({ \bar u'   u }){M}  z^\lambda    
   \int_{\Omega} d\alpha d \beta  \,   \,  z_1(\beta, \alpha,t)  \, \cos (\alpha (rz)/2 )
\,  . 
 \
 \label{Dnuc}
\end{align}

      On the lattice, choosing $z=z_3$, we can arrange $({\cal P} z)=0$, i.e. ${\cal P}_3=0$,  by taking $p_1$ and $p_2$ with 
      opposite components in $z$-direction, namely $p= (E_1, {\bf p}_{1T}, P)$ and $p'= (E_2, {\bf p}_{2T}, -P)$. 
      Introducing the relevant Ioffe time $\nu_D \equiv -(rz)/2 \Rightarrow  Pz_3$, we deal with the ITD 
       \begin{align}
  &  {\cal I}_D (\nu_D, t)
 =  \int_{-1}^1 d\alpha \, e^{i \alpha \nu_D } \, 
    {D} \left (\alpha,t\right )
\,  . 
 \
 \label{MD}
\end{align} 

However,  if we choose  $\lambda=0$,   we get \mbox{$r^0=E_1-E_2$} as the accompanying factor.
It vanishes for purely longitudinal momenta $p= (E, {\bf 0}_T, P)$,  \mbox{$p'= (E_2, {\bf 0}_T, -P)$,}
and remains rather small when one takes non-equal transverse momenta ${\bf p}_{1T}, {\bf p}_{2T} $. 

Another choice is to  take $\lambda=3$. In this case, we have $\sim z_3$ contamination 
  \begin{align}
 & \frac1{i}  \langle (E_2, {\bf p}_{2T}, -P) |  {\cal O}_-^3 (z)|(E_1, {\bf p}_{1T}, P)  \rangle 
  \nn & 
 = 2P
  \int_{-1}^1 d\alpha  \, \sin(\nu_D \alpha) \,  D(\alpha,t) \ \nn & 
\,  +  z^{(3)} M^2 
  \int_{-1}^1 d\alpha  \,\cos (\nu_D \alpha)  \,  Z_1(\alpha,t) \ ,
 \
 \label{Dpion}
\end{align} 
where the ``$Z$-term'' function $Z_1(\alpha,t)$ is even in  $\alpha$. 
Multiplying by $z_\lambda =z_3$, we have 
  \begin{align}
 & {i}  \langle(E_2, {\bf p}_{2T},-P)  | z_\lambda {\cal O}_-^\lambda (z)|(E_1, {\bf p}_{1T}, P)  \rangle 
  \nn & 
 = \nu_D
  \int_{-1}^1 d\alpha  \, \sin(\nu_D \alpha) \,  D(\alpha,t) \ \nn & 
\,  +  \frac{\nu_D^2}{4P^2}
  \int_{-1}^1 d\alpha  \,\cos (\nu_D \alpha)  \,  Z(\alpha,t)
  \nn &=
  \nu_D {\cal I}_D (\nu_D,t)+ \frac{\nu_D^2}{4P^2} {\cal I}_Z(\nu_D,t)
 \
 \label{DZpion}
\end{align} 
As we see, for a fixed $\nu$, the contamination term decreases with $P$. In principle, one may try to extract 
${\cal I}_D (\nu_D,t)$ by fitting the $P$-dependence of the matrix element. 

\section{Summary}

    In the present paper, we have outlined  the  approach of lattice extraction of GPDs based on a combined use
    of the double distributions formalism and pseudo-PDF framework. 
    The use of DDs guarantees that GPDs obtained from them have the required  polynomiality 
    property that imposes a non-trivial correlation between $x$- and $\xi$-dependences of GPDs. 
    
We have introduced  Ioffe-time distributions   writing these directly in terms of DDs, and generalized them 
 onto  correlators  off the light cone.  
 An important advantage of using DDs is  that the $D$-term appears then as an independent quantity
 rather than a non-separable  part of GPDs $H$ and $E$. 
 
We  have discussed the relation of the $\xi$-dependence of GPDs with the width of the $\alpha$-profiles of the corresponding 
DDs, and proposed  strategies for fitting lattice-extracted pseudo-distributions by DDs.  
The approach described in the present paper is  already used  in ongoing  lattice extractions of GPDs
by HadStruc collaboration.

\acknowledgements

I thank J. Karpie and  K. Orginos    for  their 
 interest in  this investigation and discussions.   
This work is supported by Jefferson Science Associates,
 LLC under  U.S. DOE Contract \mbox{ \#DE-AC05-06OR23177} 
 and by U.S. DOE Grant \#DE-FG02-97ER41028.


\end{document}